\documentclass[12pt]{iopart}

\pdfoutput=1
\usepackage{iopams}  
\usepackage{graphicx}
\begin{document}

\title[Multiple quantum oscillations]{Multiple quantum oscillation frequencies in $\mathrm{YBa_{2}Cu_{3}O_{6+\delta}}$ and bilayer splitting}

\author{David Garcia-Aldea and Sudip Chakravarty}
\address{Department of Physics and Astronomy, University of California Los Angeles \\Los Angeles, CA 90095-1547, USA}
\ead{sudip@physics.ucla.edu}
\begin{abstract}
Experiments have revealed multiple quantum oscillation frequencies in underdoped high temperature superconductor
 $\mathrm{YBa_{2}Cu_{3}O_{6+\delta}}$, corresponding to approximately $10\%$ doping, which contain $\mathrm{CuO}$ bilayers in the unit cell. These unit cells are further coupled along the $c$-axis by a tunneling matrix element. A model of the energy dispersion that has its roots in the previously determined electronic structure, combined with two-fold commensurate density waves, reveals multiple electron and hole pockets. To the extent that quasiparticles of the reconstructed Fermi surface have finite residues, however small, the formation of  Landau levels are the cause of these oscillations and the bilayer splitting and warping of the electronic dispersion along the direction perpendicular to the CuO-planes are firm consequences. We explore this possibility in detail and find overall consistency with experiments.  An important conclusion is that bilayer splitting is  considerably renormalized from the value obtained from band structure calculations.  It would be extremely interesting to perform these experiments  for higher values of doping. We roughly expect the splitting of the frequencies to increase with doping, but the full picture may be more complex because the density wave order parameter is also expected to decrease with doping, vanishing around the middle of the superconducting dome.
 
 \end{abstract}

\maketitle

\section{Introduction}
The surprising quantum oscillations (QO) in both hole~\cite{Doiron-Leyraud:2007,Jaudet:2008,LeBoeuf:2007,Sebastian:2008,Yelland:2008,Sebastian:2009,Audouard:2009,Vignolle:2008,Bangura:2008,Singleton:2009,Rourke:2009} and electron doped cuprates~\cite{Helm:2009} have raised an important question concerning the ground state of high temperature superconductors~\cite{Chakravarty:2008}. Applied magnetic field between $35-85$T  has been argued  to quench the superconducting fluctuations, at least to a large degree, revealing the normal state. This is not surprising in electron doped  $\mathrm{Nd_{2-x}Ce_{x}CuO_{4}}$ (NCCO), where the upper critical field $H_{c2}$ is less than $10\; \textrm{T}$, but is somewhat surprising in hole doped cuprates where $H_{c2}$ is extrapolated to be of order  $100\; \textrm{T}$ or greater~\cite{li:2007}.  One of the striking recent findings is the observation of multiple quantum oscillation frequencies~\cite{Audouard:2009,Ramshaw:2010}.  To understand QO in hole doped  $\mathrm{YBa_{2}Cu_{3}O_{6+\delta}}$ (YBCO) and stoichiometric  $\mathrm{YBa_{2}Cu_{4}O_{8}}$ (Y124), we shall  follow a reasoning based on broken translational symmetry with perhaps an unconventional order parameter, $d_{x^{2}-y^{2}}$-density wave (DDW)~\cite{Chakravarty:2001}. The observed multiple frequencies should not only impose constraints on the theoretical models but also interpretation of experiments, as discussed recently~\cite{Ramshaw:2010}. Superficially similar  results can be obtained within a mean field approximation using a spin density wave (SDW) theory, but we favour singlet DDW for numerous reasons discussed elsewhere~\cite{Dimov:2008}. More importantly, the quasiparticles of a singlet DDW have charge-$e$, spin-$1/2$, and a $g$-factor renormalized by residual Fermi liquid corrections. In the simplest  treatment given here, we set $g=2$. This characterization of the quasiparticles is consistent with a very recent measurement and its precise analysis~\cite{Ramshaw:2010} and perhaps eliminates any triplet order parameter, such as SDW or triplet DDW.

Not only do the experiments involving multiple quantum oscillation frequencies indicate formation of Landau levels signifying finite quasiparticle residues even in underdoped cuprates, but also indicate coherent electron motion along the direction perpendicular to the CuO-plane. A bilayer Hamiltonian corresponding to YBCO was first written down in a paper in which  an interlayer tunneling theory of superconductivity was proposed~\cite{Chakravarty:1993}. This Hamiltonian was subsequently derived from a downfolding process in a band structure calculation~\cite{Andersen:1995}. As long as the fermionic quasiparticles exist as excitations of the normal ground state, it is impossible to deny the existence of bilayer splitting, which results from the  superposition of the electronic states of the layers within a bilayer block. For each value of momentum, there is  a bonding and an antibonding state that are split in energy. In the original context~\cite{Chakravarty:1993} it was argued that only in a superconducting state such a coherent linear superposition is possible. However, it is clear that the only requirement is the existence of a finite quasiparticle residue. An important effect discussed earlier~\cite{Dimov:2008} is that the phase of the DDW order parameter of the two layers within a bilayer block make a large difference. Even though the bilayer splitting can be substantial, the splitting of the Fermi surface areas for the out-of-phase case can be very small as compared to the in-phase case. We shall focus on these two alternatives amongst other considerations.

For many years it has been argued that the normal state of high temperature superconductors is incoherent, especially in the underdoped regime. Here we shall focus on very low temperatures, where a sharp statement can be made. The view that the normal state is a non-Fermi liquid appears to be at variance with the striking QO experiments mentioned above. We look for consistency with recent experiments~\cite{Audouard:2009,Ramshaw:2010} involving multiple frequencies, emphasizing of course the general aspects of a mean field theory. A further motivation is a measurement in a tilted magnetic field~\cite{Sebastian:2010}, where inconsistency of a scenario in which observed multiple frequencies arise from bilayer-split pockets is pointed out. The idea of probing QO with a tilted field   is important, but our theoretical analyses are not in agreement with those presented in Ref.~\cite{Sebastian:2010}. 

We emphasize a  commensurate density wave order as the cause of Fermi surface reconstruction as revealed in quantum oscillation measurements, although some evidence for incommensuration does exist~\cite{Sebastian:2008}. The  pressing questions can hopefully be addressed  in a simpler setting: Why should the Fermi liquid picture be valid for the normal state? Is the motion along the direction perpendicular to the CuO-planes ($c$-axis) coherent? Why do other experimental probes of the electronic structure paint a very different picture of the fermiology? In reality, no direct evidence for any kind of long range density wave order exists in the regime of interest to the QO measurements. Fluctuating order does not solve this dilemma, especially because the QO measurements require very large correlation lengths and nearly static  order. The simplest possible explanation of  the main  aspects of the measurements call for long range order. Moreover, there are strong arguments from detailed fits to the measurements that the relatively high magnetic field is not the root of these observations~\cite{Rourke:2009}, beyond the obvious effect of suppressing superconductivity. Indeed, previous NMR measurements in  $\mathrm{YBa_{2}Cu_{4}O_{8}}$ up to at least  $23.2\; \textrm{T}$ have shown no signatures of field induced order. Yet the QO measurements for this stoichiometric material are clear and unambiguous. Of course, NMR measurements~\cite{Zheng:1999} in higher fields of the order of $45$T would be interesting. Given these larger issues and many others, it is not particularly attractive to focus on details such as incommensurate versus commensurate order. In any case, it was shown previously~\cite{Dimov:2008} that within mean field theory it is quite simple to incorporate  incommensurate order with very little change of the big picture; to go beyond mean field theory is quite difficult and is not particularly fruitful without a sufficiently strong motivation.  An important point with regard to DDW is that it is hidden from most common probes and its existence perhaps could have gone unnoticed. 

The present manuscript is organized as follows: in Section 2 we set up the effective Hamiltonian and discuss bilayer splitting in Section 3. In Section 4 we discuss our results in a perpendicular magnetic field and in Section 5 those in a tilted field. In Section 6  we discuss how variation of parameters provide contrasting evidence of the out-of-phase versus in-phase DDW order. In Sec. 7 we discuss in detail  the temperature dependences and the oscillation  magnitudes of both the magnetization and the specific heat  within the Lifshitz-Kosevich-Luttinger  formula but with Dingle factors reflecting vortex scattering rate in the mixed state.  Section 8 contains remarks regarding  unresolved puzzles.

\section{Hamiltonian}
\begin{figure}[htb]
\label{fig:unitcell}
\begin{center}
\includegraphics[width=8.5 cm]{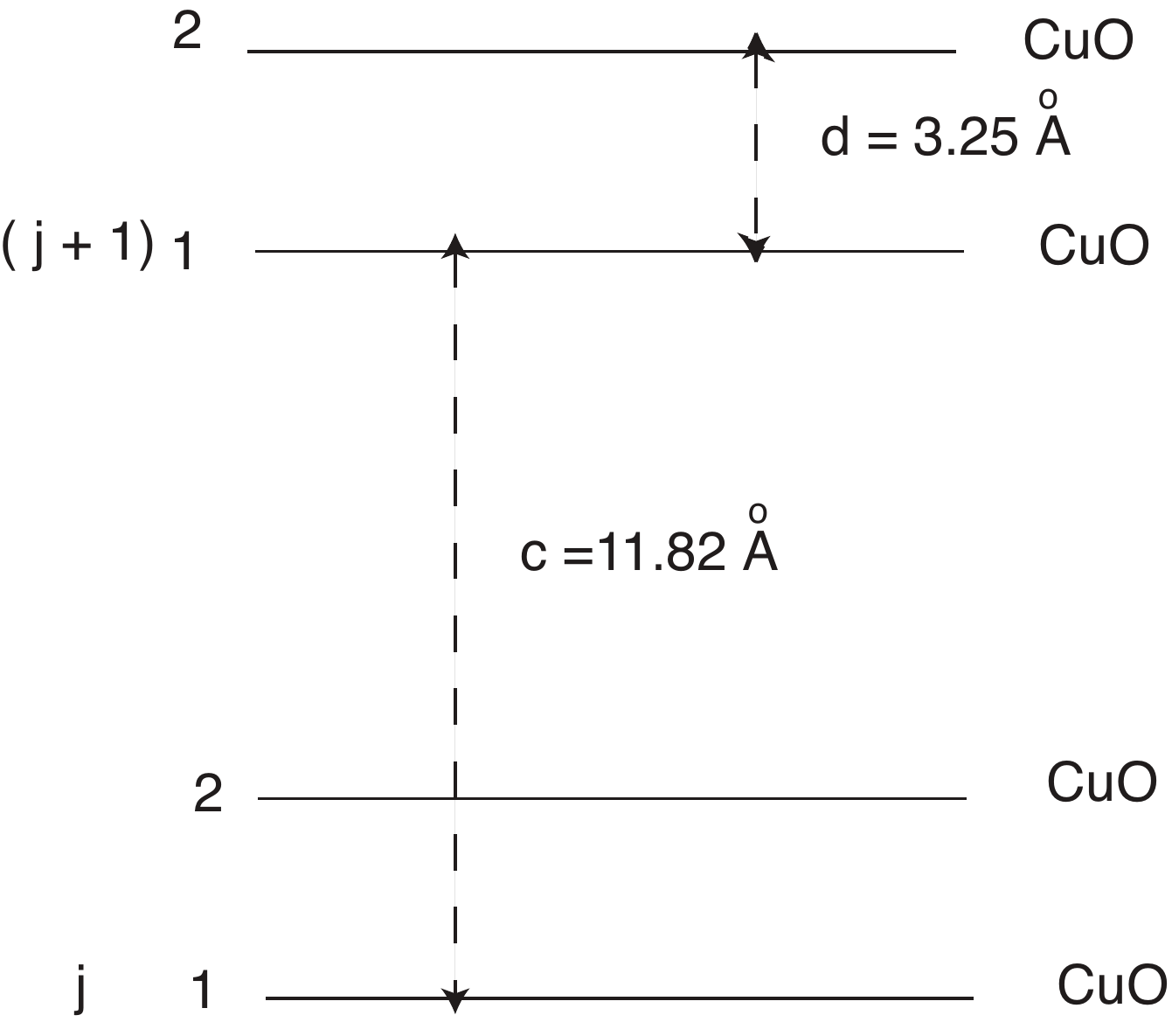}
\end{center}
\caption{Bilayer structure of $\mathrm{YBa_{2}Cu_{3}O_{6+\delta}}$. Each unit cell contains a bilayer CuO block. Note that tunneling matrix elements within a bilayer unit and between the nearest neighbour planes between the unit cells are kept; other matrix elements are exponentially smaller.}
\end{figure}

We consider a tight-binding Hamiltonian, $H_{0}$, which captures correctly the bilayer splitting and the matrix elements between the unit cells; see Figure~\ref{fig:unitcell}:
\begin{eqnarray}
H_{0}&=&\sum_{j,{\bf k}}\sum_{n=1}^{2}\epsilon({\bf k})c^{\dagger}_{n,j}({\bf k})c_{n,j}({\bf k})-\sum_{j,{\bf k}}t_{\perp}({\bf k})c^{\dagger}_{1,j}({\bf k})c_{2,j}({\bf k})+h. c.\nonumber \\&-&t_{c}\sum_{j,{\bf k}}c_{1,j+1}^{\dagger}({\bf k})c_{2,j}({\bf k})+h.c.
\end{eqnarray}
The fermion annihilation operator $c_{n,j}({\bf k})$ depends on the bilayer index $n$ within the unit cell and the index $j$ refers to the unit cell. The spin indices are suppressed and ${\bf k}=(k_{x},k_{y})$ is a two-dimensional vector. Only the hopping matrix element, $t_{c}$, between the nearest neighbor planes of two adjacent unit cells are kept, as the tunnelling matrix elements to further neighbour planes are considerably smaller. For simplicity $t_{c}$ is assumed to be momentum independent, as very little is known about its precise form. This assumption will have little effect on our analysis. The bilyaer matrix element corresponding to $\mathrm{YBa_{2}Cu_{3}O_{6+\delta}}$ is~\cite{Chakravarty:1993}
\begin{equation}
t_{\perp}({\bf k})=\frac{t_{\perp}}{4}\left[\cos(k_{x}a)-\cos(k_{y}a)\right]^{2},
\end{equation}
where $a$ is the in-plane lattice constant, ignoring slight orthorhombicity. $H_{0}$ can be further simplified by the canonical transformation~\cite{Klemm:1991},
\begin{equation}
c_{n}({\bf k},k_{z})= \frac{1}{\sqrt{M}}\sum_{j}c_{n,j}({\bf k})e^{ik_{z}[jc+(n-1)d]} e^{\mp i\phi(k_{z})/2},
\end{equation}
which diagonalizes it  in the momentum space. Note the additional phase factors, $e^{-i\phi(k_{z})}$ for $n=1$ and  $e^{+i\phi(k_{z})}$ for $n=2$. The choice of the phase $\phi(k_{z})=-k_{z}d$ preserves the fermion anticommutaion rules and results in an energy spectrum that is periodic in $2\pi /c$, which preserves the periodicity of the conventional unit cell.  The $2\times 2$ bilayer block is still not diagonal and must be diagonalized further to obtain the quasiparticle dispersion. Note that $t_{c}$ is a matrix element between the nearest neighbour planes of the two bilayer blocks and will be chosen to be an adjustable parameter. The canonical transformation leads to 
\begin{equation}
\fl H_{0}=\sum_{k_{z},{\bf k}}\left\{\sum_{n=1}^{2}\epsilon_{\bf k}c^{\dagger}_{n}(k_{z,}{\bf k})c_{n}(k_{z},{\bf k})-[t_{\perp}({\bf k})+t_{c}e^{-ik_{z}c}]c^{\dagger}_{1}(k_{z},{\bf k})c_{2}(k_{z},{\bf k})+h.c.\right\}
\label{eq:Eq4}
\end{equation}
We use a common band structure~\cite{Andersen:1995,Pavarini:2001},
\begin{eqnarray}\label{dispersion}
\fl\epsilon_{\bf k}=-2t(\cos{k_x}a+\cos{k_y}a)+4t'\cos{k_x}a\cos{k_ya}-2t''(\cos{2k_x}a+\cos{2k_y}a).
\end{eqnarray}
with $t'=0.32 t$ and $t''=\frac{1}{2}t'$, but $t$ is chosen to be  $t\approx 0.1 eV $. This renormalized value  of $t$, as compared  to the band structure value of $0.38 eV$, seems to be phenomenologically more appropriate in the underdoped regime of interest to us. However, the specific results pertaining to the ground state at $T=0$ are independent of the magnitude of $t$;  even if we had chosen $t=0.38 eV$, the  results would have been the same provided the remaining parameters are chosen proportionately. This is no longer be true when  we consider the $T\ne 0$ properties discussed in Sec. 6. We shall first choose  $t_{\perp}=0.05 t$ and $t_{c}= 0.013 t$; these parameters are  expected to be highly renormalized in the underdoped regime. Even when bilayer splitting is clearly observed in angle resolved photoemission spectroscopy (ARPES) in heavily overdoped  $\mathrm{Bi_{2}Sr_{2}CaCu_{2}O_{8+\delta}}$ (Bi2212)~\cite{Feng:2001}, the actual magnitude of $t_{\perp}$ is severely overestimated by the band structure calculations, $300$ meV, as opposed to the observed $88$ meV. In the underdoped regime, the band structure value is likely to be  more unreliable because of strong correlation effects. In Sec. 6 we shall see how the variation of $t_{\perp}$ affects the principal conclusions.

\begin{figure}[htb]
\label{fig:warp}
\begin{center}
\includegraphics[width=8.5 cm]{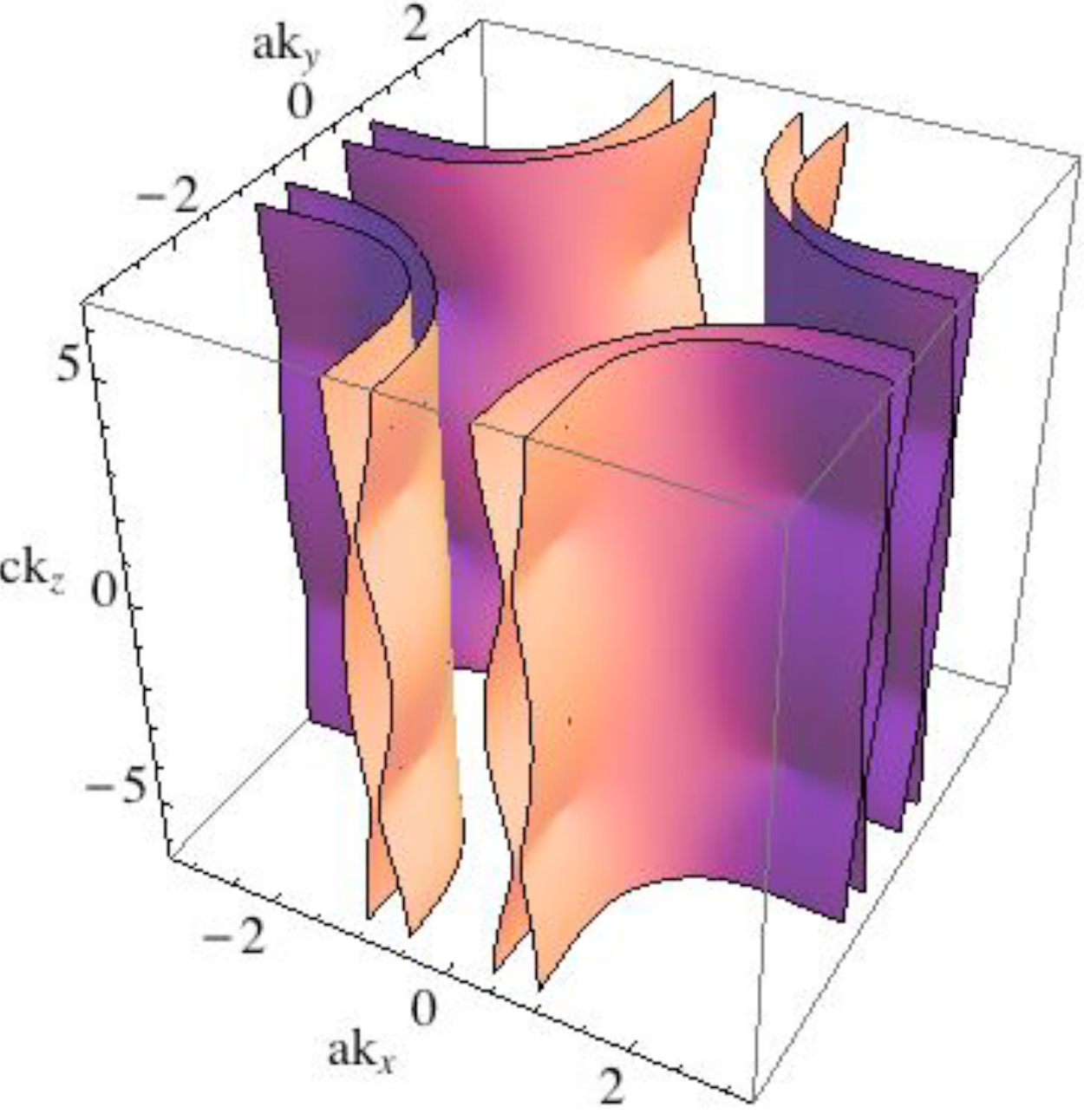}
\end{center}
\caption{A greatly exaggerated illustration of warping of the bilayer split bands plotted in the extended zone $-2\pi\le k_{z}c\le 2 \pi$, $-\pi\le k_{x}a\le \pi$, and $-\pi\le k_{y}a\le \pi$. }
\end{figure}

The $2\times 2$ Hamiltonian has the eigenvalues 
\begin{equation}
\lambda_{\pm}(k_{z},{\bf k})=\epsilon_{\bf k}\pm \sqrt{t_{c}^{2}+t_{\perp}({\bf k})^{2}+2 t_{c} t_{\perp}({\bf k}) \cos k_{z}c} \; .
\end{equation}
It is interesting to note that with our choice of the phase of the fermion operators the distance between the layers in a bilayer block, $d$, does not appear explicitly in the spectrum, only implicitly in the magnitude of the hopping matrix elements. The above result is very different from the conventional warping of layered materials that contain only one electronically active plane per unit cell, which leads to a dispersion $\epsilon_{\bf k}- 2 t_{c}\cos k_{z}c$. 
It is clear that the two bilayer split bands are warped differently as a function of $k_{z}$, as seen in Figure~\ref{fig:warp}.
As long as  $t_{c}$ is nonzero, the splitting at the nodal locations $(\pi/2a,\pi/2a)$ is non-zero. We now fold the Brillouin zone to the reduced Brillouin zone (RBZ)  bounded by $k_{y}\pm k_{x}=\pm \pi/a$, corresponding to the two-fold commensurate singlet DDW order parameter and augment $H_{0}$ by $H'$:
\begin{equation}
H'=\sum_{k_{z},{\bf k} \in RBZ}\left[i \, W_{\mathbf{k}} c_{1}(k_{z},\mathbf{k})^{\dagger} c_{1}(k_{z},{\bf
k}+{\bf Q})+h.c.\right]+(1\to 2),
\end{equation}
where ${\bf Q} =(\pi/a, \pi/a)$ and the DDW gap $W_{\bf k}$ is real and is given by
\begin{equation}
W_{\bf k}=\frac{W_{0}}{2}(\cos k_{x}a - \cos k_{y}a).
\end{equation}
Note that the DDW order parameters are chosen to be in-phase for the layers.  To reproduce the experimental frequencies we  require a somewhat large value of $W_{0}=0.85 t$ within our mean field approximation; in Sec. 6 we shall consider a small variation of this parameter.

The singlet DDW condensate is defined by~\cite{Nayak:2000}
\begin{equation}\label{corder_k}
\langle{c}^{\dagger}_{n',\sigma'}({\bf k'},k_{z})c_{n,\sigma}({\bf k},k_{z})\rangle
= iW_{\bf k}\,\delta_{\sigma',\sigma}\delta_{n',n}
\delta_{{\bf k'},{\bf k+Q}},
\end{equation}
Note that it involves $\delta_{\sigma',\sigma}$ for spin indices. This is the reason why  the spin indices can be conveniently suppressed. This is a particle-hole condensate that breaks the following symmetries:  translation by a lattice spacing,  time reversal, parity, and a rotation by $\pi/2$, while the product of any two are preserved. The order parameter corresponds to angular momentum $\ell =2$. Since there is no exchange symmetry between a particle and a hole, the orbital wave function does not determine the spin wave function. Therefore, there is also a corresponding triplet DDW, which consists of a staggered pattern of circulating spin currents~\cite{Nayak:2000}, as opposed to a staggered pattern circulating charge currents. In the present work, we shall consider only the singlet DDW order and make only brief remarks regarding the triplet DDW at the very end. The staggering is determined by the wave vector $\bf Q$.
\section{Bilayer splitting}
\begin{figure}[htb]
\begin{center}
\includegraphics[width=8.5 cm]{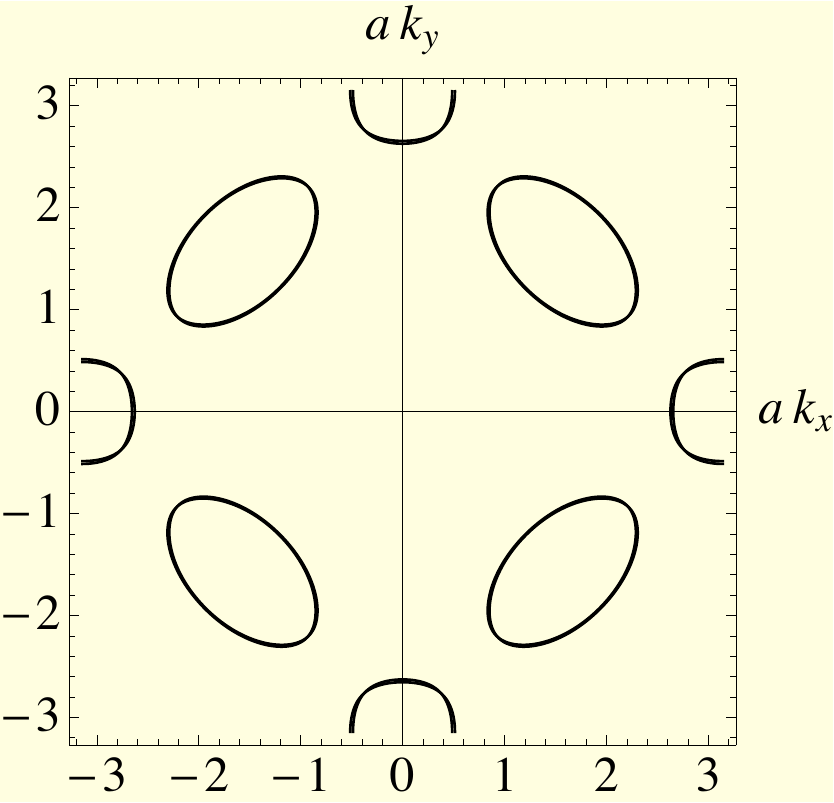}
\end{center}
\caption{Bilayer splitting of the Fermi surfaces for the out-of-phase DDW order parameter. The cut is at $k_{z}=0$. For clarity the contours are plotted in the extended zone.}
\label{fig:split2}
\end{figure}

The combined Hamiltonian $H_{0}+H'$ can be written in terms of the four component spinor $\Psi^{\dagger}({\bf k}, k_{z})=\{c^{\dagger}_{1}({\bf k},k_{z}), c^{\dagger}_{1}({\bf k+Q},k_{z}), c^{\dagger}_{2}({\bf k},k_{z}), c^{\dagger}_{2}({\bf k+Q},k_{z})\}$, suppressing once again the spin indices, which is irrelevant for a singlet DDW order parameter. In terms of this spinor the combined Hamiltonian is
\begin{equation}
\mathbb{H}=\sum_{{\bf k}\in RBZ, k_{z}} \Psi^{\dagger}({\bf k},k_{z}) \mathbb{A}  \Psi({\bf k},k_{z})
\end{equation}
where
\begin{equation}
\fl\mathbb{A} = \left(\begin{array}{cccc}\epsilon_{\bf k} & iW_{\bf k} & -t_{\perp} ({\bf k})-t_{c}e^{-ick_{z}}& 0 \\- iW_{\bf k} & \epsilon_{{\bf k}+{\bf Q}}  & 0 & -t_{\perp} ({\bf k}) -t_{c}e^{-ick_{z}}\\ -t_{\perp} ({\bf k})-t_{c}e^{ick_{z}}& 0 & \epsilon_{\bf k} &  iW_{\bf k} \\0 &  -t_{\perp} ({\bf k})-t_{c}e^{ick_{z}} & - iW_{\bf k}  & \epsilon_{{\bf k}+{\bf Q}} \end{array}\right).
 \label{eq:h}
\end{equation}
Note that the DDW order parameters in the two $2\times 2$ diagonal blocks are in phase. The {\em in-phase} DDW order parameter corresponds to ``ferromagnetically'' aligned  staggered circulating currents in the layers  within a bilayer block. 
The four eigenvalues of the matrix $\mathbb{A}$ are
\begin{equation}
\fl\lambda_{1\pm}^{s}({\bf k})=\frac{\epsilon_{\bf k}+\epsilon_{\bf k+Q}}{2}\pm\left | \sqrt{\left(\frac{\epsilon_{\bf k}-\epsilon_{\bf k+Q}}{2}\right)^{2}+W_{\bf k}^{2}}-\sqrt{t_{c}^{2}+t_{\perp}({\bf k})^{2}+2 t_{c} t_{\perp}({\bf k}) \cos k_{z}c}\right|,
\end{equation}
and
\begin{equation}
\fl\lambda_{2\pm}^{s}({\bf k})=\frac{\epsilon_{\bf k}+\epsilon_{\bf k+Q}}{2}\pm\left ( \sqrt{\left(\frac{\epsilon_{\bf k}-\epsilon_{\bf k+Q}}{2}\right)^{2}+W_{\bf k}^{2}}+\sqrt{t_{c}^{2}+t_{\perp}({\bf k})^{2}+2 t_{c} t_{\perp}({\bf k}) \cos k_{z}c}\right)
\end{equation}
\begin{figure}[htb]
\begin{center}
\includegraphics[width=8.5 cm]{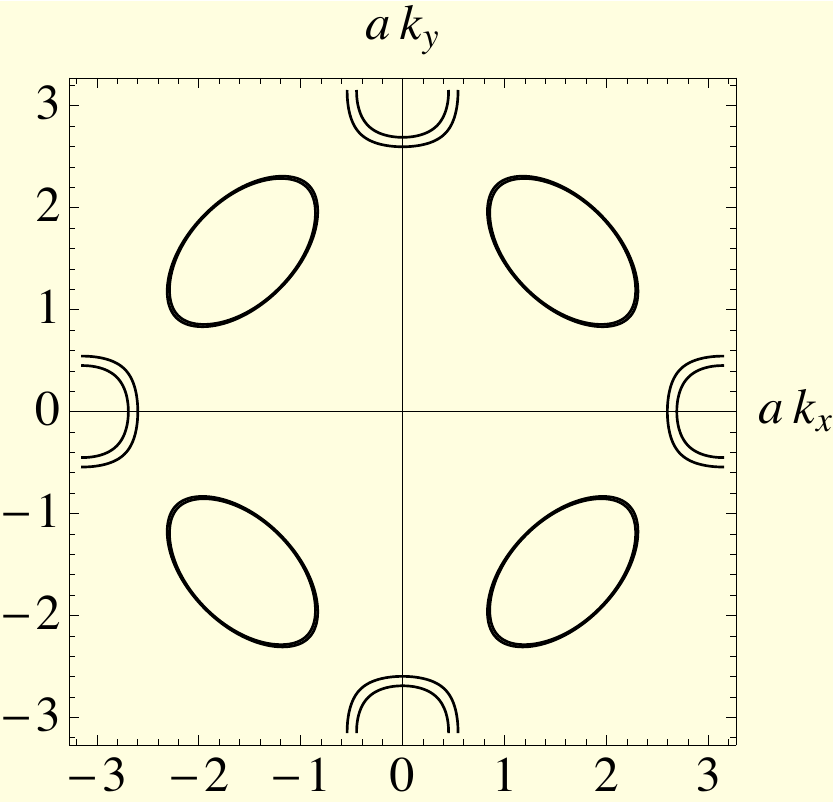}
\label{fig:split1}
\end{center}
\caption{Bilayer splitting of the Fermi surfaces for the  in-phase DDW order parameter. The cut is at $k_{z}=0$ where the splitting is maximal. For clarity the contours are plotted in the extended zone.}
\end{figure}
For a particle-hole condensate,  when measured from the chemical potential, the spectra are 
\begin{eqnarray}
E_{1\pm}^{s}({\bf k})&=&\lambda_{1,\pm}^{s}({\bf k})-\mu, \\
E_{2\pm}^{s}({\bf k})&=&\lambda_{2,\pm}^{s}({\bf k})-\mu.
\end{eqnarray}
because both $\epsilon({\bf k})$ and $\epsilon({\bf k}+{\bf Q})$ are equally shifted by $\mu$.
If, on the other hand, the staggered circulating currents are ``antiferromagnetically''  aligned  within a bilayer block~\cite{Dimov:2008}, that is, $iW_{\bf k}$ is replaced by  $-iW_{\bf k}$ in the lower $2\times 2$ diagonal block ({\em out-of-phase}),  the corresponding eigenvalues are
 \begin{equation}
\fl\lambda_{1\pm}^{a}({\bf k})=\frac{\epsilon_{\bf k}+\epsilon_{\bf k+Q}}{2}\pm\left\{W_{\bf k}^{2}+ \left(\left|\frac{\epsilon_{\bf k}-\epsilon_{\bf k+Q}}{2}\right|-\sqrt{t_{c}^{2}+t_{\perp}({\bf k})^{2}+2 t_{c} t_{\perp}({\bf k}) \cos k_{z}c}\right)^{2}\right\}^{1/2},
\end{equation}
and
 \begin{equation}
\fl\lambda_{2\pm}^{a}({\bf k})=\frac{\epsilon_{\bf k}+\epsilon_{\bf k+Q}}{2}\pm\left\{W_{\bf k}^{2}+ \left(\left|\frac{\epsilon_{\bf k}-\epsilon_{\bf k+Q}}{2}\right|+\sqrt{t_{c}^{2}+t_{\perp}({\bf k})^{2}+2 t_{c} t_{\perp}({\bf k}) \cos k_{z}c}\right)^{2}\right\}^{1/2}.
\end{equation}
Once again, measured from $\mu$ , we have 
\begin{eqnarray}
E_{1\pm}^{a}({\bf k})&=&\lambda_{1,\pm}^{a}({\bf k})-\mu, \\
E_{2\pm}^{a}({\bf k})&=&\lambda_{2,\pm}^{a}({\bf k})-\mu.
\end{eqnarray}
The contour plots for the Fermi surfaces corresponding to $\lambda_{1\pm}^{s}$ and $\lambda_{2\pm}^{s}$ for $k_{z}=0$ are shown in Figure~\ref{fig:split1}. It is clear that while the electron pockets are observably split, the splitting of the hole pockets is much smaller. The chemical potential, $\mu= -0.78t $, was adjusted to yield approximately $10.3 \%$ hole doping.
For identical set of parameters, the splitting for the out-of-phase eigenvalues,  $\lambda_{1\pm}^{a}$ and 
$\lambda_{2\pm}^{a}$,   is considerably smaller, as shown in Figure~\ref{fig:split2}. It would be incorrect, however, to infer that the splitting is exactly zero; see Table~\ref{table2} below. Note that the absolute value of $t$ does not change the frequencies because $\epsilon _{F}\left( \mathbf{k;}\alpha t,\alpha t^{\prime },\alpha t^{\prime
\prime },\alpha t_{\perp },\alpha t_{c},\alpha W_{0}\right) =\alpha \epsilon
_{F}\left( \mathbf{k;}t,t^{\prime },t^{\prime \prime },t_{\perp
},t_{c},W_{0}\right)$, as long as we also let $\mu \to \alpha \mu$.
\section{Magnetic field perpendicular to the CuO-plane: Onsager relation}
Quantum oscillation frequencies can be obtained from the extremal areas, ${\cal A}(\epsilon_{F})$, of the Fermi surface perpendicular to the applied magnetic field~\cite{Shoenberg:1984}. The Onsager relation for the frequency $F$ is
\begin{equation}
F = \frac{\hbar c}{2\pi e}{\cal A}(\epsilon_{F}).
\end{equation}
Of course, this formula presupposes that the quasiclassical approximation is valid and there are no significant magnetic breakdown effects.

Given the electronic structure, the doping dependence can be obtained from noting that there are two hole pockets within the RBZ and one electron pocket. These are further split by the bilayer coupling and warped by the $k_{z}$ dependence. Taking into account two spin directions, the doping fraction of a given electron pocket corresponding to the bilayer bonding band $b$, $x^{b}_{e}$ is 
\begin{equation}
x^{b}_{e}=2\frac{2a^{2}c}{(2\pi)^{3}}\int_{0}^{\pi/a} dk_{x}\int_{0}^{\pi/a}dk_{y}\int_{-\pi/c}^{\pi/c}dk_{z} \; \theta\left(\mu - \epsilon^{b}({\bf k},k_{z})\right).
\end{equation}
There is an identical expression for the antibonding contribution $x^{a}_{e}$. Similarly the two hole pockets contribute an amount $x^{b}_{h}$ given by
\begin{equation}
x^{b}_{h}=2\frac{2a^{2}c}{(2\pi)^{3}}\int_{0}^{\pi/a} dk_{x}\int_{0}^{\pi/a}dk_{y}\int_{-\pi/c}^{\pi/c}dk_{z} \; \theta\left(\epsilon^{b}({\bf k},k_{z})-\mu\right),
\end{equation}
with an identical antibonding contribution $x^{a}_{h}$. The total hole doping per CuO-plane  is then
\begin{equation}
x_{h}=\frac{1}{2}(x^{b}_{h}+x^{a}_{h}-x^{b}_{e}-x^{a}_{e}).
\end{equation}

The frequencies for the in-phase order parameter are given in Table~\ref{table1}. The parameters were chosen, but not particularly optimized, to be similar to the observed frequencies~\cite{Audouard:2009} $540\pm 15$T, $630\pm 40$T, $450\pm 15$T and $1130\pm 20$T. Out of 4 theoretically predicted frequencies corresponding to the electron pocket only 3 are observed. The fourth observed frequency at $1130$T could correspond to the hole pocket that is split very little. Alternately, it may also be a harmonic. It has been puzzle for some time~\cite{Chakravarty:2008b,Morinari:2009,Jia:2009,Rourke:2009,Harrison:2007,Harrison:2009} as to why the hole pocket frequencies have such weak or non-existent  signatures in quantum oscillation measurements.
\begin{table}
\caption{\label{table1}Bilayer split frequencies for the in-phase DDW order. Here $t_{\perp}=0.05 t$, $t_{c}=0.013 t$ and doping is approximately $\sim10.3\%$. The band parameters are given in the text. The electron pocket is labelled as e-pocket and the hole pocket as the h-pocket.}
\begin{indented}
\item[]\begin{tabular}{@{}llll}
\br
e-pocket ($k_{z}=0$)&e-pocket ($k_{z}=\pi/c$)& h-pocket ($k_{z}=0$)& h-pocket ($k_{z}=\pi/c$)\\
\mr
711 T& 659 T& 1051 T&1032 T\\
480 T & 534 T& 997 T& 1015 T\\
\br
\end{tabular}
\end{indented}
\end{table}

In contrast, the out-of-phase frequencies (Table~\ref{table2}) do not resemble the experimental observations~\cite{Audouard:2009}. We provide an alternate picture in Sec. 6 based on the experiment in Ref.~\cite{Ramshaw:2010}.  Within the mean field approximation adopted here, it is not possible to distinguish between the in-phase and the out-of-phase cases~\cite{Dimov:2008} as far as the electronic energy is concerned. For this one would need a detailed microscopic Hamiltonian. This is outside the scope of the present investigation. We therefore rely on  experiments to distinguish between the two cases.
\begin{table}
\caption{\label{table2}Bilayer split frequencies for the out-of--phase DDW order. The parameters are the same as in Table~\ref{table1}.}
\begin{indented}
\item[]\begin{tabular}{@{}llll}
\br
e-pocket ($k_{z}=0$)&e-pocket ($k_{z}=\pi/c$)& h-pocket ($k_{z}=0$)& h-pocket ($k_{z}=\pi/c$)\\
\mr
617 T & 609 T& 1044 T&1031 T\\
573 T & 585 T& 1002 T& 1016 T\\
\br
\end{tabular}
\end{indented}
\end{table}

\section{Tilted magnetic field}

\begin{figure}[htb]
\begin{center}
\includegraphics[width=8.5 cm]{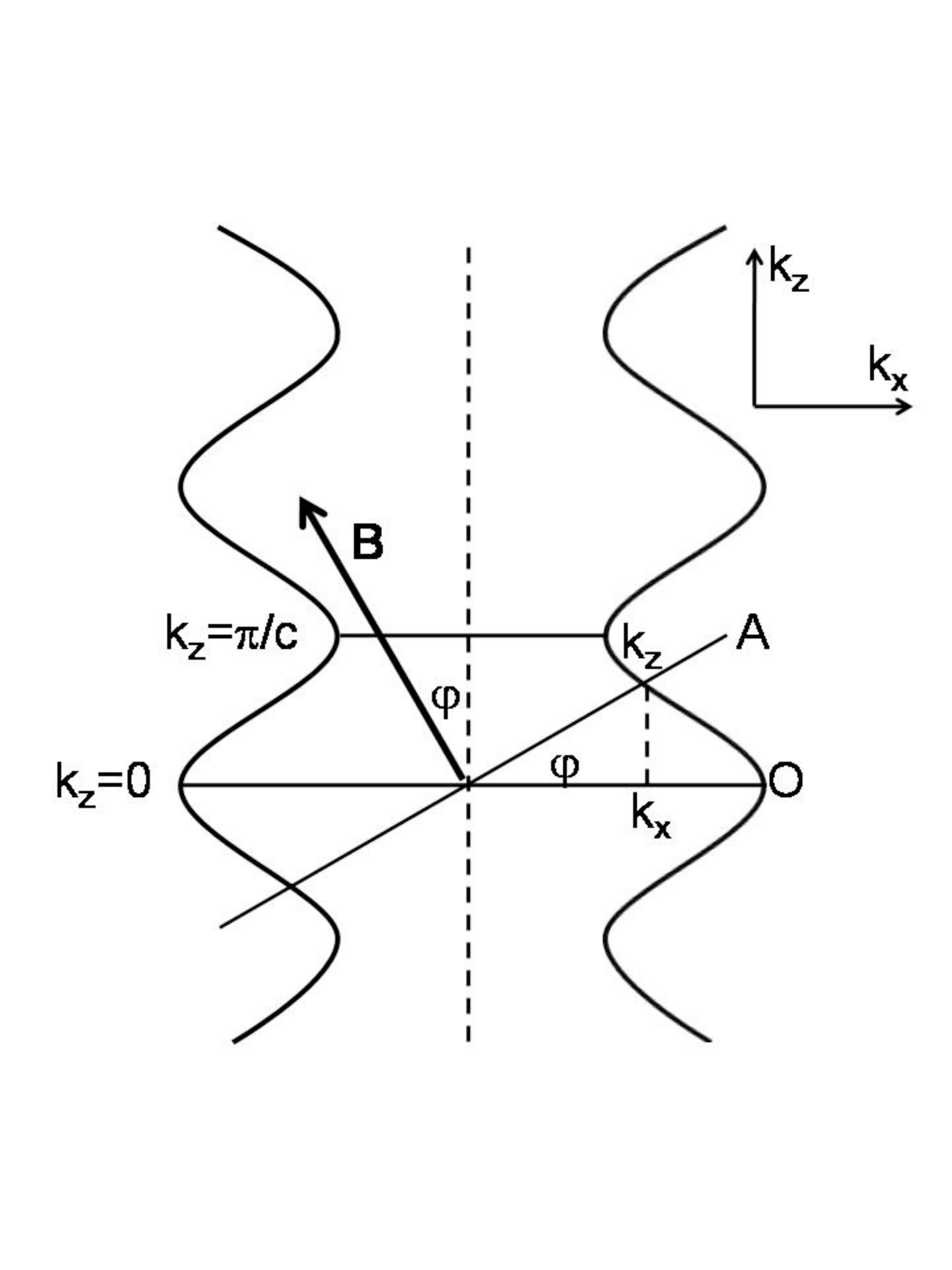}
\end{center}
\caption{A cut of the Fermi surface of a pocket by a plane $k_{y}=-\pi/a$. $\bf B$ is the applied magnetic tilted at an angle $\varphi$. }
\label{fig:FigAngle1}
\end{figure}

\begin{figure}[htb]
\begin{center}
\includegraphics[width=8.5 cm]{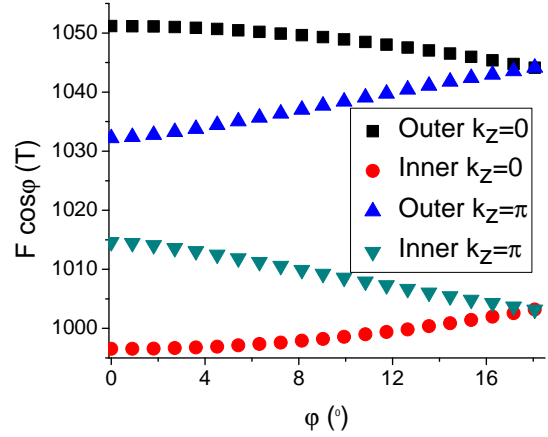}
\end{center}
\caption{Hole pocket frequencies as a function of the tilt angle $\varphi$.}
\label{fig:Freq1}
\end{figure}

\begin{figure}[htb]
\begin{center}
\includegraphics[width=8.5 cm]{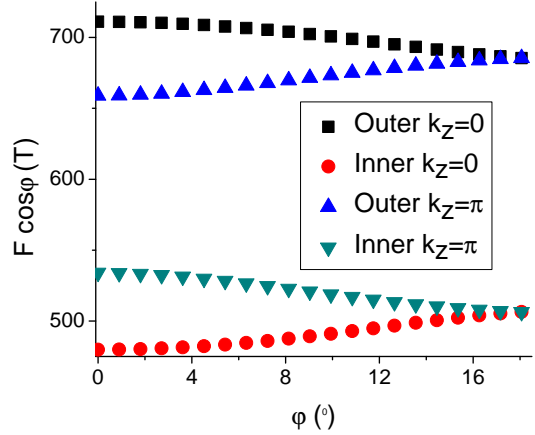}
\end{center}
\caption{Electron pocket frequencies as a function of the tilt angle $\varphi$.}
\label{fig:Freq2}
\end{figure}

\begin{figure}[htb]
\begin{center}
\includegraphics[width=8.5 cm]{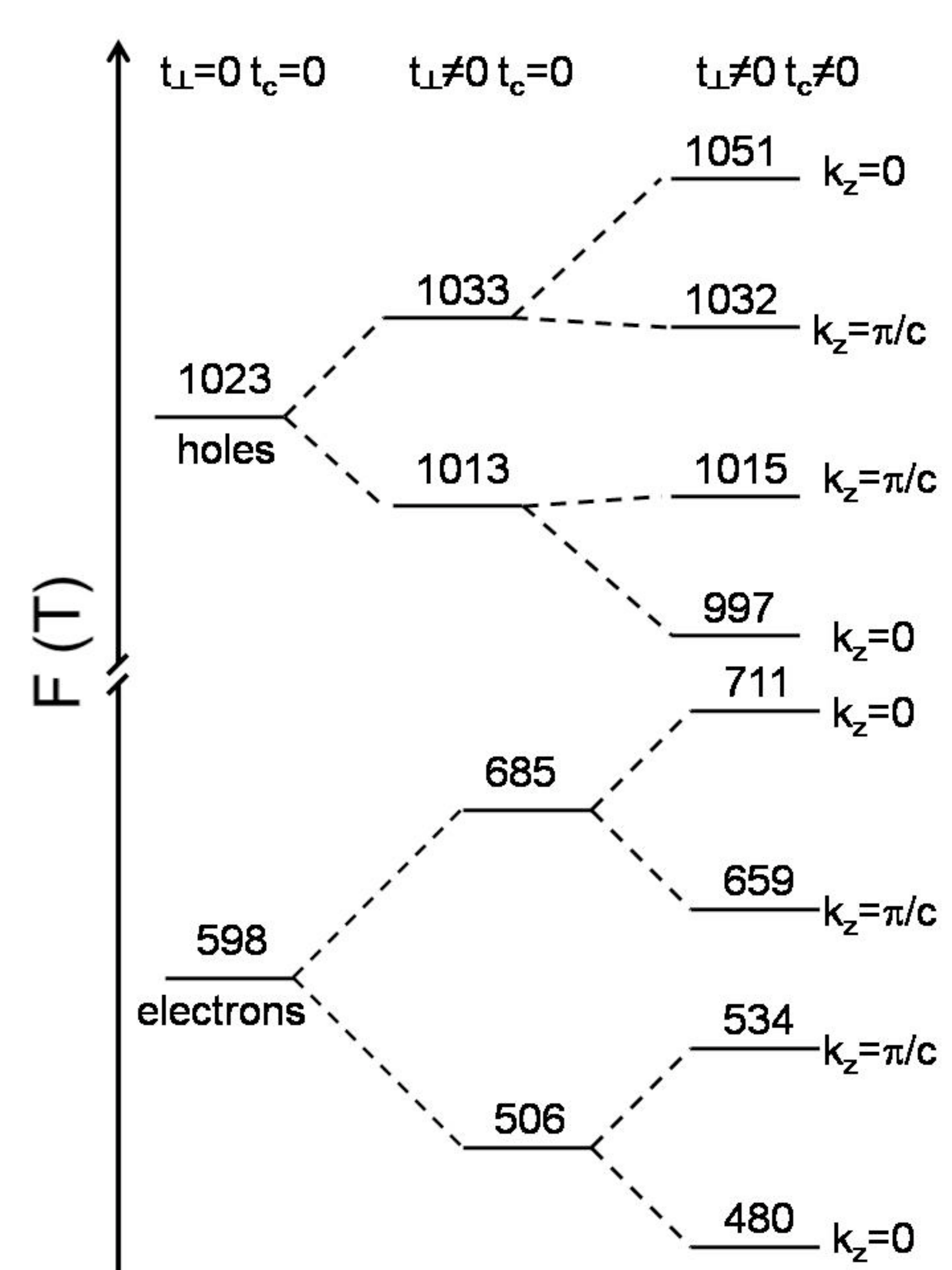}
\end{center}
\caption{The hierarchy of frequency splitting for in-phase DDW order parameter with the magnetic field normal to the CuO-plane, as we sequentially turn on $t_{\perp}$ and $t_{c}$, not to scale.}
\label{fig:Freqs1}
\end{figure}

In this section we calculate the effect of tilted magnetic field on quantum oscillations~\cite{Yamaji:1989}. In Figure~\ref{fig:FigAngle1} we show a cut of the Fermi surface with the plane $k_{y}=-\pi/a$. The intersection with the plane $A$ is given by
\begin{equation}
\epsilon _{F}\left( k_{x},k_{y},k_{x}\tan \varphi \right) =\mu 
\end{equation}
If the Fermi surface does not depend $k_{z}$, the area ${\cal A}_{O}$ will be constant for all planes perpendicular to $k_{z}=0$, and the area in the plane A will be given by  ${\cal A}_{A}={\cal A}_{O}/\cos \varphi$ with a constant value of ${\cal A}_{O}$.
However, given the dependence on $k_{z}$, the area is 
\begin{equation}
{\cal A}_{A}\left( \varphi \right) \cos \varphi =\int_{-\pi /a}^{-\pi
/a}dk_{x}\int_{-\pi /a}^{-\pi /a}dk_{y}\theta \left[ \epsilon _{F}\left(
k_{x},k_{y},k_{x}\tan \varphi \right) -\mu \right],
\end{equation}
which can be computed numerically. The above result corresponds to $k_{z}=0$. More generally, when the plane $O$ is situated at an arbitrary value of $k_{z}$, we get
\begin{equation}
\label{eq:tilt}
\fl {\cal A}_{A}\left( k_{z},\varphi \right) \cos \varphi =\int_{-\pi /a}^{-\pi
/a}dk_{x}\int_{-\pi /a}^{-\pi /a}dk_{y}\theta \left[ \epsilon _{F}\left(
k_{x},k_{y},k_{z}+k_{x}\tan \varphi \right) -\mu \right] .
\end{equation}
Note that Equation~\ref{eq:tilt} is valid for angles $\varphi\le \varphi_{max}$ such that 
\begin{equation}
\tan \varphi_{max}=\frac{\pi/c}{\pi/a}=\frac{a}{c}
\end{equation}
Beyond this maximum angle, there are discontinuous jumps, and we do not attempt to treat this case. 
For hole pockets the frequencies are summarized in Figure~\ref{fig:Freq1}.
Similarly for the electron pockets the frequencies are shown in Figure~\ref{fig:Freq2}.
Note that the vertical scales are different in Figures~\ref{fig:Freq1} and~\ref{fig:Freq2}. It is also interesting to 
note how the frequencies are sequentially split as we first turn on $t_{\perp}$ and then $t_{c}$, which is shown in Figure~\ref{fig:Freqs1} for the magnetic field in the direction $k_{z}$.
\begin{figure}[htb]
\begin{center}
\includegraphics[width=8.5 cm]{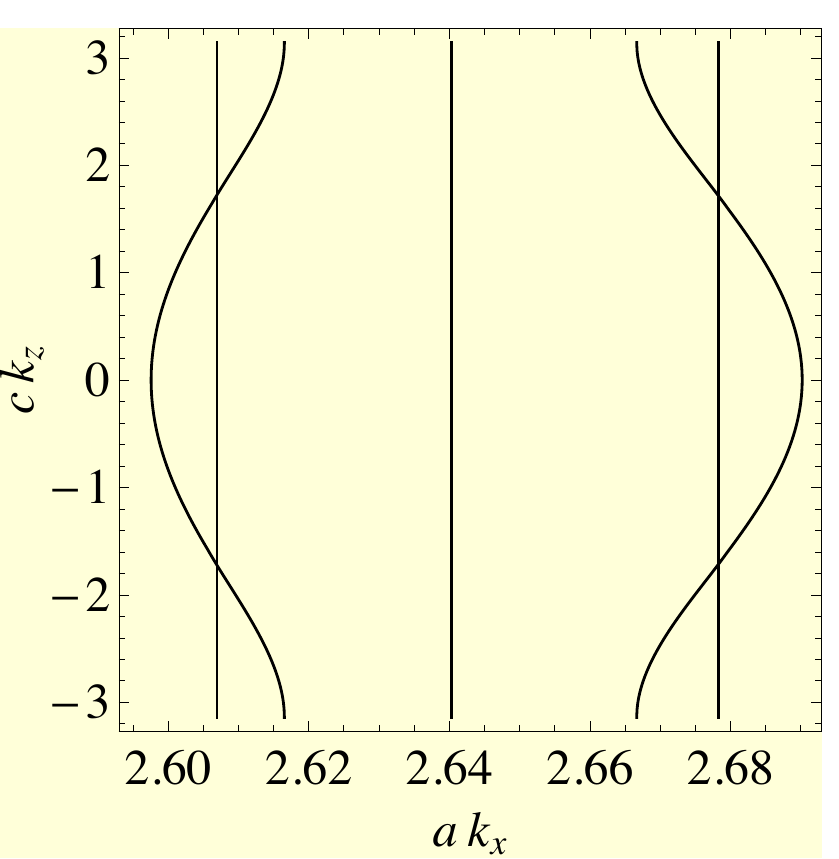}
\end{center}
\caption{Warping of the bilayer split electron pockets.}
\label{fig:SplitWarp}
\end{figure}

In Figure~\ref{fig:SplitWarp} we illustrate the warping along
$k_{z}$ for an electron pocket. We represent a cut for $
k_{x}=0$ that yields $k_{y}a$ around $2.64$. 
The vertical line at the center
corresponds to a model where  $t_{\perp}=t_{c}=0$. 
The two vertical lines at the left and the right of the figure show the
splitting  when  $t_{\perp}\ne 0$. The line on the left corresponds to the outer pocket and the line on
the right to the inner pocket. As we turn on  $t_{c}$, the  warping is seen as two curved lines. It is clear
that the warping has opposite sense for the outer and the inner pockets.
The bilayer splitting can be  seen from the displacement of the left line by  $0.0334$ from the central line, while the right line is displaced by $0.0379$ in the oposite direction. The splitting induced by
$t_{\perp}$ is therefore not symmetric. To calculate the warping we can compute distance between the lines
at $k_{z}=0$, obtaining $0.00937$ for the outer pocket and $0.0119$ for the
inner one. At $k_{z}=\pm \pi /c$, $ 0.00958$ is the displacement  for the outer pocket and $0.01158$ is the displacement 
for the inner one. {\em These numbers encode two important facts: first, the warping is
different  for the inner and the outer pockets,  and, second, it cannot be
modeled with a simple cosine dependence.}

\section{Variation of parameters}
Here we vary the parameters to see how the results change. The focus is the difference between the out-of-phase and the in-phase DDW order parameters. We have already seen that there is a qualitative distinction between them. However, given the recent measurements~\cite{Ramshaw:2010}, we would like to see if one or the other can be made more consistent with these experiments. We stress that the phenomenological nature of our work precludes us  from fitting parameters with certainty, nor is it our intention. We only look for some qualitative insights. However, since in this section we shall be computing the oscillatory part of the thermodynamic potential, as a function of temperature and magnetic field, not just the frequencies, a good estimate of the leading tight-binding matrix element $t$ is necessary for materials relevant for quantum oscillation experiments.  Since there are no reliable ARPES for YBCO, the next best we can do is to rely on the recent tight-binding fit to the measured ARPES in Y124~\cite{Kondo:2009}, a system in which good quantum oscillations have been observed.  Except for $t$, the ratios of the remaining band parameters to $t$ are not very different from the band structure results given below Equation~\ref{eq:Eq4}. Thus, we simply take over the value of $t$ determined from ARPES in Y124, which is $t=0.154\; \textrm{eV}$ (the average of the fit to the bonding and the antibonding bands). Additionally, we would like to see if one can tolerate a much larger value of bilayer matrix element, $t_{\perp}$, as compared to the earlier section and still find consistency with experiments. We shall see that this is indeed possible, but only for the out-of-phase DDW order parameter. 

\begin{figure}[htb]
\begin{center}
\includegraphics[width=8.5 cm]{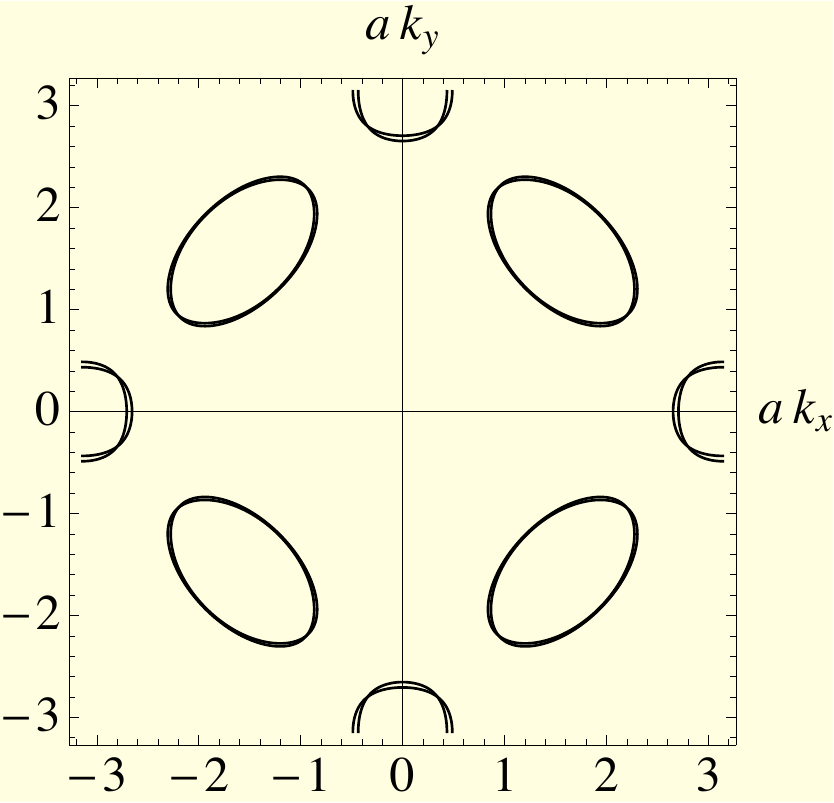}
\end{center}
\caption{Bilayer splitting for the out-of-phase DDW order parameter. The cut is at $k_{z}=0$. For clarity the contours are plotted in the extended zone. Here $t_{\perp}=0.12 t$ and doping is $\approx 10.7\%$. The band structure  parameters, $t'$ and $t''$,  are unchanged, $t'=0.32 t$ and $t''=0.5 t$, but $W_{0}=0.9 t$, where $t=0.154\;\textrm{eV}$.}
\label{fig:split3}
\end{figure}

In this section we keep all the band structure parameters fixed, including $t_{c}$, but  more than double the bilayer matrix element to $t_{\perp}=0.12 t = 0.0185\; \textrm{eV}$, resulting in a splitting of $37\; \textrm{meV}$, which is reasonable compared to the overdoped Bi2212, where it is measured to be $88\; \textrm{meV}$; one expects  renormalization with underdoping. To keep the doping level more or less fixed ($\approx 10.7\%$), we set $\mu = -0.775 t$ and $W_{0}=0.9 t$. The resulting oscillation frequencies are shown in Table~\ref{table3}, and the Fermi surfaces at $k_{z}=0$ are plotted in Fig.~\ref{fig:split3}. 
\begin{figure}[htb]
\begin{center}
\includegraphics[width=8.5 cm]{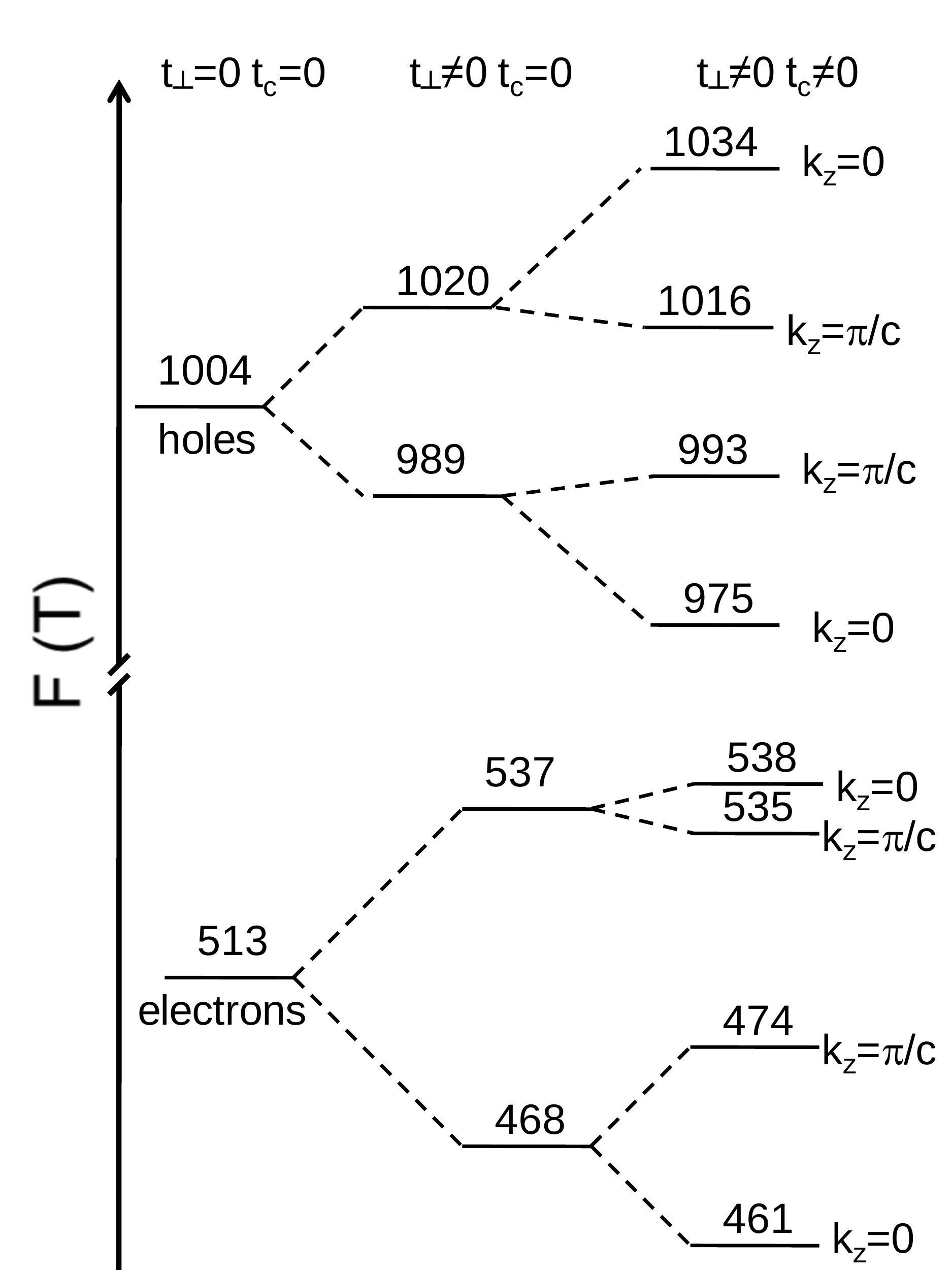}
\end{center}
\caption{The hierarchy of frequency splitting for out-of-phase DDW order parameter with the magnetic field normal to the CuO-plane, as we sequentially turn on $t_{\perp}$ and $t_{c}$, not to scale.}
\label{fig:Freqs2}
\end{figure}
\begin{table}
\caption{\label{table3}Bilayer split frequencies for the out-of-phase DDW order parameter. Here $t_{\perp}=0.12 t$ and doping is $\approx 10.7\%$. The band structure  parameters, $t'$ and $t''$,  are unchanged, $t'=0.32 t$ and $t''=0.5 t$, but $W_{0}=0.9 t$, where $t=0.154\;\textrm{eV}$.}
\begin{indented}
\item[]\begin{tabular}{@{}llll}
\br
e-pocket ($k_{z}=0$)&e-pocket ($k_{z}=\pi/c$)& h-pocket ($k_{z}=0$)& h-pocket ($k_{z}=\pi/c$)\\
\mr
538 T & 535 T & 1034 T& 1015 T\\
461 T & 474 T & 975 T& 993 T\\
\br
\end{tabular}
\end{indented}
\end{table}
The two groups of electron pocket frequencies are close to each other and so are the two groups of  hole pocket frequencies despite much larger bilayer splitting. The warping of the outer electron pocket is only $3\; \textrm{T}$ and that of the inner pocket is $13\; \textrm{T}$.  It is even possible to tolerate larger $t_{\perp}$, but we have not explored it further. It is again useful to examine the frequency diagram. This is shown in Fig.~\ref{fig:Freqs2} and is quite different from Fig.~\ref{fig:Freqs1}.

In contrast, for the same set of parameters, as above, the in-phase DDW results in frequencies that are no longer close to the recent experiments~\cite{Ramshaw:2010} as shown in Table~\ref{table4}. The warping of the outer electron pocket is $51\; \textrm{T}$ and that of the inner electron pocket is $57\; \textrm{T}$.
\begin{table}
\caption{\label{table4}Bilayer split frequencies for the in-phase DDW order parameter. Here $t_{\perp}=0.12 t$ and doping is $\approx 10.7\%$. The band structure  parameters are unchanged, $t'=0.32 t$ and $t''=0.5 t$, but $W_{0}=0.9 t$, where $t=0.154\;\textrm{eV}$. Compare with Table~\ref{table3}.}
\begin{indented}
\item[]\begin{tabular}{@{}llll}
\br
e-pocket ($k_{z}=0$)&e-pocket ($k_{z}=\pi/c$)& h-pocket ($k_{z}=0$)& h-pocket ($k_{z}=\pi/c$)\\
\mr
750 T & 699 T & 1046 T& 1022  T\\
250 T & 307 T & 965 T& 988 T\\
\br
\end{tabular}
\end{indented}
\end{table}

\subsection{Cyclotron masses and the second derivatives of the extremal areas}
In the following section we shall need the cyclotron masses and the second derivatives of the 
extremal areas. 
These are calculated numerically and 
are sumarized in Table~\ref{table5} and Table~\ref{table6}.
\begin{table}[tbp]
\caption{Cyclotron masses in units of the free electron mass  for the out-of-phase DDW order
parameter.}
\label{table5}
\begin{indented}
\item[]\begin{tabular}{@{}llll}
\br
e-pocket ($k_{z}=0$)&e-pocket ($k_{z}=\pi/c$)& h-pocket ($k_{z}=0$)& h-pocket ($k_{z}=\pi/c$)\\
\mr
1.67 & 1.64 & 0.99 & 0.97 \\
1.47 & 1.49 & 0.93 & 0.95 \\
\br
\end{tabular}
\end{indented}
\end{table}
The second derivatives of the extremal areas with respect to
$k_{z}$ are more difficult to calculate. We  fit the
areas near $k_{z}=0$ and $k_{z}=\pi $  by 
a fourth order polynomial with only even terms (odd terms are zero within numerical precision)
\begin{eqnarray}
\mathcal{A} &\approx &\mathcal{A}_{0}+\frac{\mathcal{A}_{2}}{2}
(k_{z}c)^{2}+\frac{\mathcal{A}_{4}}{24}(k_{z}c)^{4} \\
\mathcal{A} &\approx &\mathcal{A}_{0}+\frac{\mathcal{A}_{2}}{2}\left[
(k_{z}c)-\pi \right] ^{2}+\frac{\mathcal{A}_{4}}{24}\left[ (k_{z}c)-\pi \right]
^{4}
\end{eqnarray}
\begin{table}[tbp]
\caption{The fitting coefficients of the extremal areas.}
\label{table6}
\begin{indented}
\item[]\begin{tabular}{@{}llll}
\br
&$\mathcal{A}_{0}$&$\mathcal{A}_{2}$&$\mathcal{A}_{4}$\\
\mr
h-pocket ($k_{z}=0$)&1.45542& -0.00804& -0.00051\\
h-pocket ($k_{z}=0$)&1.37238& 0.00776& -0.00488\\
e-pocket ($k_{z}=0$)&0.75696&-0.00119& -0.00305\\
e-pocket ($k_{z}=0$)&0.64928&0.00902& -0.02421\\
h-pocket ($k_{z}=\pi/c$)&1.43000&0.04116& -1.16330\\
h-pocket ($k_{z}=\pi/c$)&1.39761&-0.04102& 1.18022\\
e-pocket ($k_{z}=\pi/c$)&0.75315&0.00269& -0.01303\\
e-pocket ($k_{z}=\pi/c$)&0.66771&-0.00862& 0.00637\\
\br
\end{tabular}
\end{indented}
\end{table}
It is interesting to note that while the cyclotron masses depend on the in-plane hopping matrix element $t$, the second derivatives of the extremal areas are independent of $t$.

\section{Oscillation amplitudes of specific heat and magnetization}
Within Fermi liquid theory Luttinger~\cite{Luttinger:1961} has shown that the thermodynamic potential is given by ($\beta=1/k_{B}T$)
\begin{equation}
\Omega = -\frac{1}{\beta}\sum_{r}\ln \left[1+e^{\beta (\mu - E_{r})}\right]
\end{equation}
where $\{E_{r}\}$ constitute the spectra of elementary excitations behaving like independent particles in a magnetic field, including Fermi liquid corrections; $r$ denotes the collection of quantum numbers: the Landau level $n$, $k_{z}$, and the spin $\sigma$. The spectra $\{E_{r}\}$ in a crystalline solid in high magnetic fields are of course not easy to calculate, especially if we have to include bilayer splitting and the DDW order  discussed above, but a rigorous answer can be given within  an asymptotic  expansion. Luttinger has shown that the problem maps onto to that solved by  Lifshitz and Kosevich (LK)~\cite{Abrikosov:1988} in which the thermodynamic potential depends on  the extremal areas of closed orbits, the derivative of the areas with respect to energy at the chemical potential, and the second derivative of the extremal areas with respect to $k_{z}$.  The beauty of this approach is that it is not necessary to know $E_{r}$ explicitly. Thus, even given the complexity of the present problem, the procedure to calculate the oscillatory part of the thermodynamic potential is straightforward. As with all asymptotic expansions, the validity of the procedure far surpasses what we may naively perceive to be the regime of validity. Thus the LK formula has stood the test of time, especially with Luttinger's Fermi liquid corrections.

For simplicity, in this section we shall consider magnetic field only in the $c$-direction.  Taking into account only the fundamental frequencies, $F_{i}$, the oscillatory part of $\Omega$ is 
\begin{equation}
\frac{\Omega }{V}\propto -H^{5/2}\sum_{i}\frac{1}{m^{*}_{i}}\left\vert \frac{%
\partial ^{2}S_{i}}{\partial k_{z}^{2}}\right\vert ^{-1/2}\psi \left(
\lambda_{i} \right) \cos \left[ \frac{2\pi F_{i}}{H}\pm \frac{\pi }{4}\right] \cos\left(\pi \frac{m^{*}_{i}}{m}\right). 
\end{equation}
The phase $\pm \pi/4$ correspond to positive or negative sign of the second derivative of the extremal area with respect to $k_{z}$.
The sum is over all extremal surfaces, and $m_{i}^{\ast }$ is the cyclotron effective mass given by ($m$ is free electron mass)
\begin{equation}
m_{i}^{\ast }=\frac{\hbar^{2}}{2\pi }\left\vert \frac{\partial S_{i}}{\partial \mu }
\right\vert , 
\end{equation}
$\left\vert \frac{\partial ^{2}S_{i}}{\partial k_{z}^{2}}\right\vert $ is
the second derivative of the area of the Fermi surface with respect to $
k_{z}$. The argument of the function 
\begin{equation}
\psi \left( \lambda_{i} \right) =\frac{\lambda_{i}}{\sinh \lambda_{i} } 
\end{equation}
 is 
 \begin{equation}
\lambda_{i} =\frac{2\pi ^{2}k_{B}T}{\hbar \omega _{ci}^{\ast }}
\end{equation}
The cyclotron frequencies are given by $\omega_{ci}^{*}=eH/m^{*}_{i}c$.
The oscillatory part of the specific heat is then
\begin{equation}
\frac{C_{V}^{osc}}{V}\propto -TH^{1/2}\sum_{i}m_{i}^{\ast }\left\vert \frac{%
\partial ^{2}S_{i}}{\partial p_{z}^{2}}\right\vert ^{-1/2}\psi ^{\prime
\prime }\left( \lambda_{i} \right) \cos \left[ \frac{2\pi F_{i}}{H}\pm \frac{\pi 
}{4}\right] \cos\left(\pi \frac{m^{*}_{i}}{m}\right),
\end{equation}
where $\psi ^{\prime \prime }\left( \lambda \right) $ is the second
derivative of $\psi \left( \lambda \right) $with respect to $\lambda$. Similarly, the leading oscillatory term of the magnetization is
\begin{equation}
\frac{M}{V}^{osc}\propto -H^{1/2}\sum_{i}\frac{F_{i}}{m_{i}^{\ast }}\left\vert 
\frac{\partial ^{2}S_{i}}{\partial p_{z}^{2}}\right\vert ^{-1/2}\psi \left(
\lambda_{i} \right) \cos \left[ \frac{2\pi F_{i}}{H}\pm \frac{\pi }{4}\right] \cos\left(\pi \frac{m^{*}_{i}}{m}\right).
\end{equation}
These results need to be supplemented by the Dingle factors that damp the oscillations due to scattering from defects or vortices in the vortex liquid state or both. We expect that the total scattering rate to be given by the combination of defect and vortex scattering rates
\begin{equation}
\frac{\hbar}{\tau}= \frac{\hbar}{\tau_{d}}+\frac{\hbar}{\tau_{v}}
\end{equation}
Moreover, these scattering rates must depend on the particular extremal area, $i$, under consideration. The calculation of the Dingle factors 
\begin{equation}
{\cal D}_{i} = e^{-\pi/\omega_{ci}^{*}\tau_{i}},
\end{equation}
especially in the mixed phase including disorder, with coexisting fluctuating $d$-wave superconducting order parameter and DDW, is a daunting task. Previously, we have shown rigorously that almost any form of conventional  disorder due to defects in a pure DDW state suppresses the electron pockets more than the hole pockets. For the vortex scattering rate, however, an approximate treatment based on a paper by Stephen~\cite{Stephen:1992} led to an interesting prediction relating the Dingle factors of electron and hole pockets (not including bilayer splitting)  in the commensurate case, which is 
\begin{equation}
\left(\frac{\hbar}{\omega_{c}\tau_{v}}\right)_{h}\approx  \sqrt{2}\left(\frac{\widetilde{m}}{m^{*}}\right)^{3/2}\left(\frac{\hbar}{\omega_{c}\tau_{v}}\right)_{e}
\end{equation}
where $\widetilde{m}$ is a characteristic scale having the dimension of mass corresponding to the {\em massless} nodal fermions of the DDW (!), and $m^{*}$ is the cyclotron mass corresponding to the electron pocket as defined above (note that the notations are different here from Ref.~\cite{Jia:2009}), which in turn is very close to the band mass defined by expanding around the bottom of the electron pocket. Although the precise numerical relation is difficult to control, it is reasonable to set 
${\cal D}_{h}= {\cal D}_{e}^{\alpha}$, with $\alpha = 1.5 - 4.5$ for phenomenological purposes; we had estimated this parameter earlier to be 4.4~\cite{Jia:2009}. 

\subsection{Specific heat and magnetization}
With the frequencies given in Table~\ref{table3}, the cyclotron masses in Table~\ref{table4}, and the second derivatives of the areas in Table~\ref{table5}, we can compute the oscillatory parts of the specific heat and the magnetization provided we can make reasonable estimates of the Dingle factors. The Dingle factor for electrons is a bit more controlled because the band mass obtained obtained by expanding around the antinodal points is quite consistent with the computed cyclotron masses. Assuming that samples have negligible disorder, we shall estimate the scattering rate of the electrons  to be given by the vortex scattering rate, which, following an  analysis of Stephen~\cite{Stephen:1992}, was found to be~\cite{Dimov:2008,Jia:2009}
\begin{equation}
\left(\frac{1}{\tau_{v}}\right)_{e}= \frac{\Delta_{0}^{2}}{\hbar}\left(1-\frac{H}{H_{c2}}\right)\sqrt{\frac{\pi}{|\mu|\hbar \omega_{c}}}
\end{equation}
where $\Delta_{0}$ is the magnitude of the $T=0$ superconducting gap, which we set to be $\approx 10\textrm{meV}$ for the relevant doping range. The cyclotron frequency $\omega_{c}=eH/m^{*}c$, with $m^{*}$ given in Table~\ref{table4}. With the present set of parameters, and with the average value of $m^{*}$, we find that
\begin{equation}
\left(\frac{1}{\tau_{v}}\right)_{e}\approx 8.5 \times 10^{12}\; \mathrm{s^{-1}}.
\end{equation}
where we used, as a typical case,  $H=40 \; \textrm{T}$ and $H_{c2}\approx 100\;  \textrm{T}$. We believe that this gives the correct order of magnitude; for the earlier set of parameters we estimated it to be $3\times 10^{12}\; \mathrm{s^{-1}}$~\cite{Dimov:2008}. The Dingle factors of the holes are more complex~\cite{Dimov:2008,Jia:2009} because it has to be estimated taking into account the nodal fermions for DDW, but it is not unreasonable to assume $\alpha\approx 2$ in the relation ${\cal D}_{h}= {\cal D}_{e}^{\alpha}$, uniformly for all electron and hole pockets. 

\begin{figure}[htb]
\begin{center}
\includegraphics[scale=0.5]{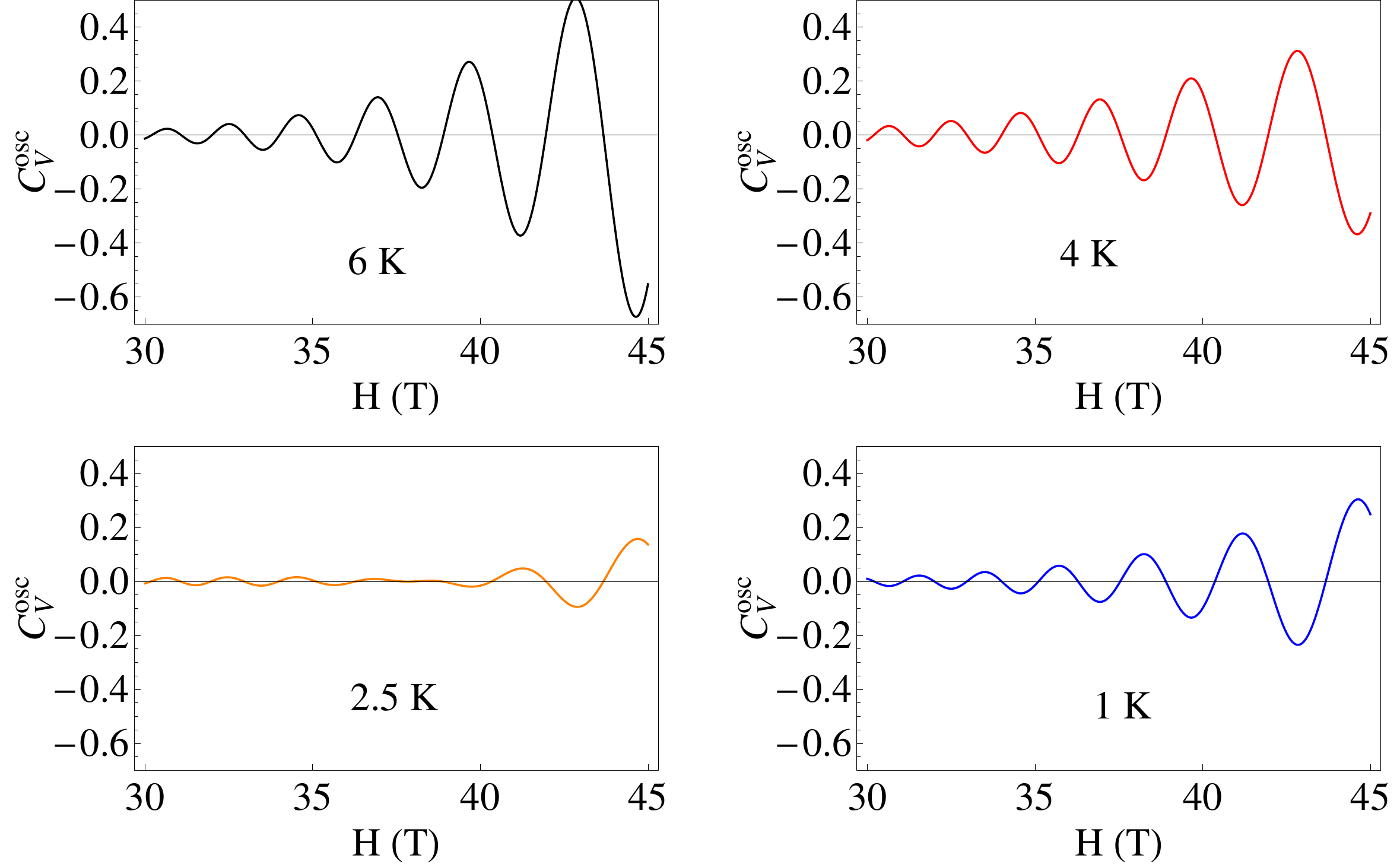}
\end{center}
\caption{Oscillatory part of the specific heat as a function of the magnetic field  at  representative temperatures. Note the phase shift by $\pi$ as the the temperature is lowered. The vertical scale is in arbitrary units.}
\label{fig:Cv1}
\end{figure}

\begin{figure}[htb]
\begin{center}
\includegraphics[scale=0.4]{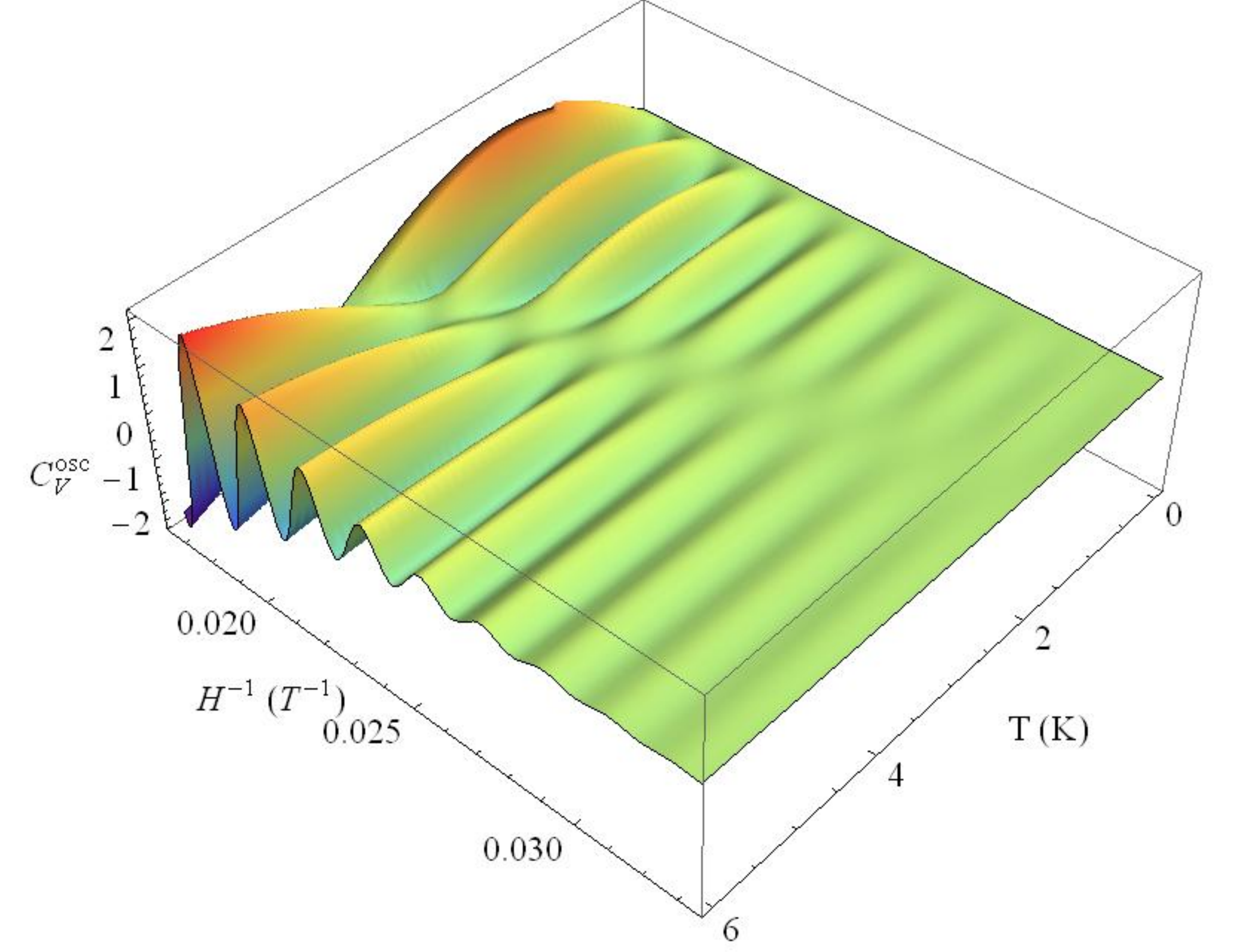}
\end{center}
\caption{Oscillatory part of the specific heat as a function of temperature and the inverse of the magnetic field. Note the presence of nodes for intermediate values magnetic field and temperature. The vertical scale is in arbitrary units}
\label{fig:Cv2}
\end{figure}

\begin{figure}[htb]
\begin{center}
\includegraphics[scale=0.4]{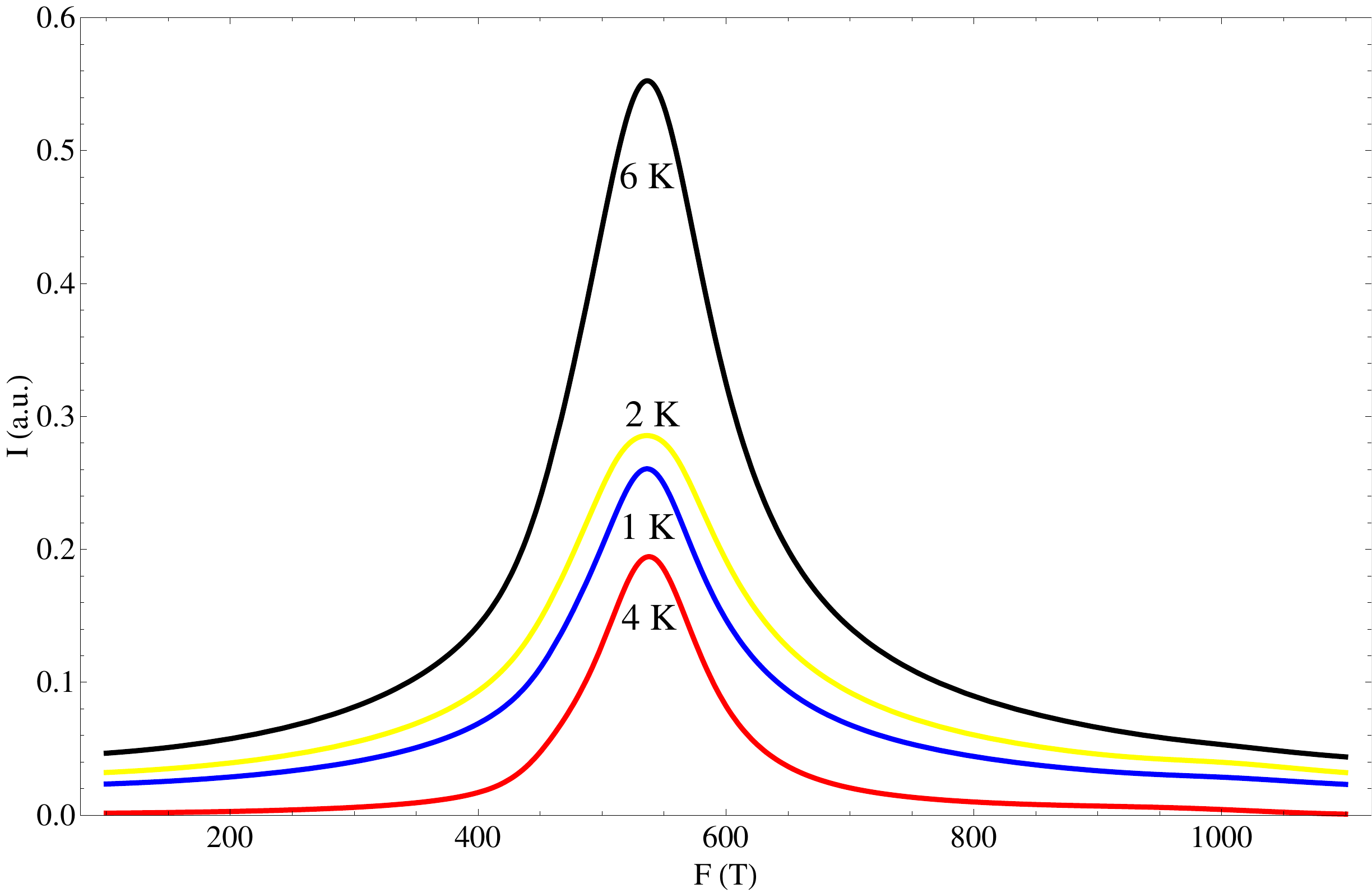}
\end{center}
\caption{Absolute value of the Fourier transform, $I$, in arbitrary units of the  oscillations of the specific heat as a function of the frequency $F$ in units of Tesla. No windowing was performed, and the field range was  $1/60 \le 1/H \le 1/20$.}
\label{fig:Cv3}
\end{figure}

The computed specific heat at four representative temperatures are shown in Fig.~\ref{fig:Cv1}. It is interesting to note that there is a $\pi$-phase shift from high to low temperatures.
The same results are visualized in a 3D-plot in Fig.~\ref{fig:Cv2}.
As a function of temperature and magnetic field the oscillations go through a node, which is also visible in Fig.~\ref{fig:Cv1}. The reason for this is the factor $\psi''(\lambda)$ in the formula for the specific heat. The Fourier transform of the oscillations in $1/H$ on the other hand shows a more complex structure for specific heat as shown in Fig.~\ref{fig:Cv3}, which, however, is very sensitive to the Dingle factor, the range of $1/H$ over which the Fourier transform is performed, and the windowing technique. The results shown here uses no windowing technique,  and the range of the magnetic field is $1/60 \le 1/H \le 1/20$.
The non-monotonic behavior of the Fourier transform in Fig.~\ref{fig:Cv3} can be understood by glancing at Fig.~\ref{fig:Cv2}. Because of the aforementioned node, the transition from $6 K$ to $4 K$ lowers the amplitude. At $2 K$ the amplitude recovers again and then finally decreases again at $1 K$. Note that only one dominant frequency is seen.

\begin{figure}[htb]
\begin{center}
\includegraphics[scale=0.4]{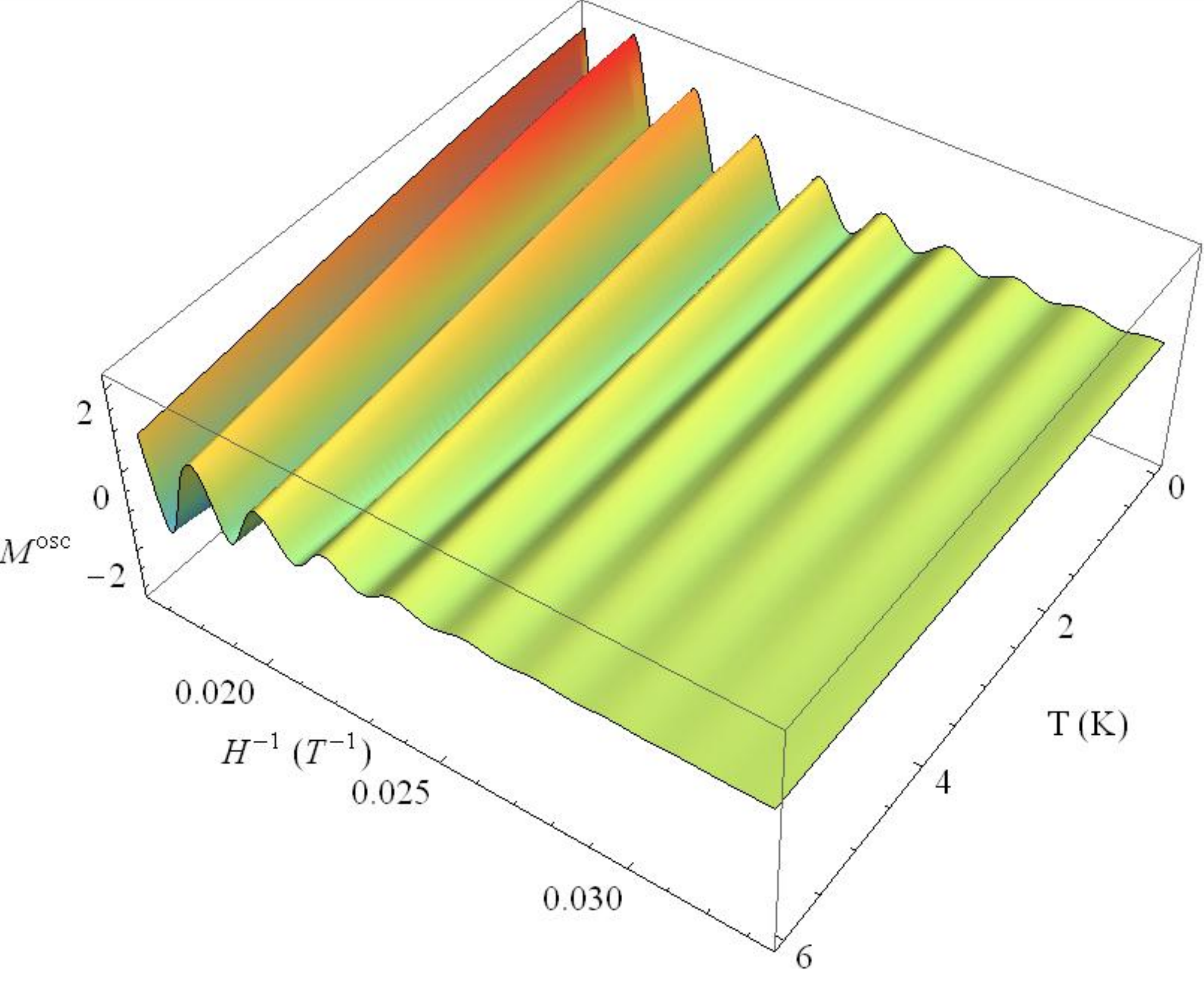}
\end{center}
\caption{Oscillatory part of the magnetization as a function of temperature and the inverse of the magnetic field.There are no nodes at intermediate temperatures as in the case of specific heat. The vertical scale is in arbitrary units}
\label{fig:M2}
\end{figure}

\begin{figure}[htb]
\begin{center}
\includegraphics[scale=0.4]{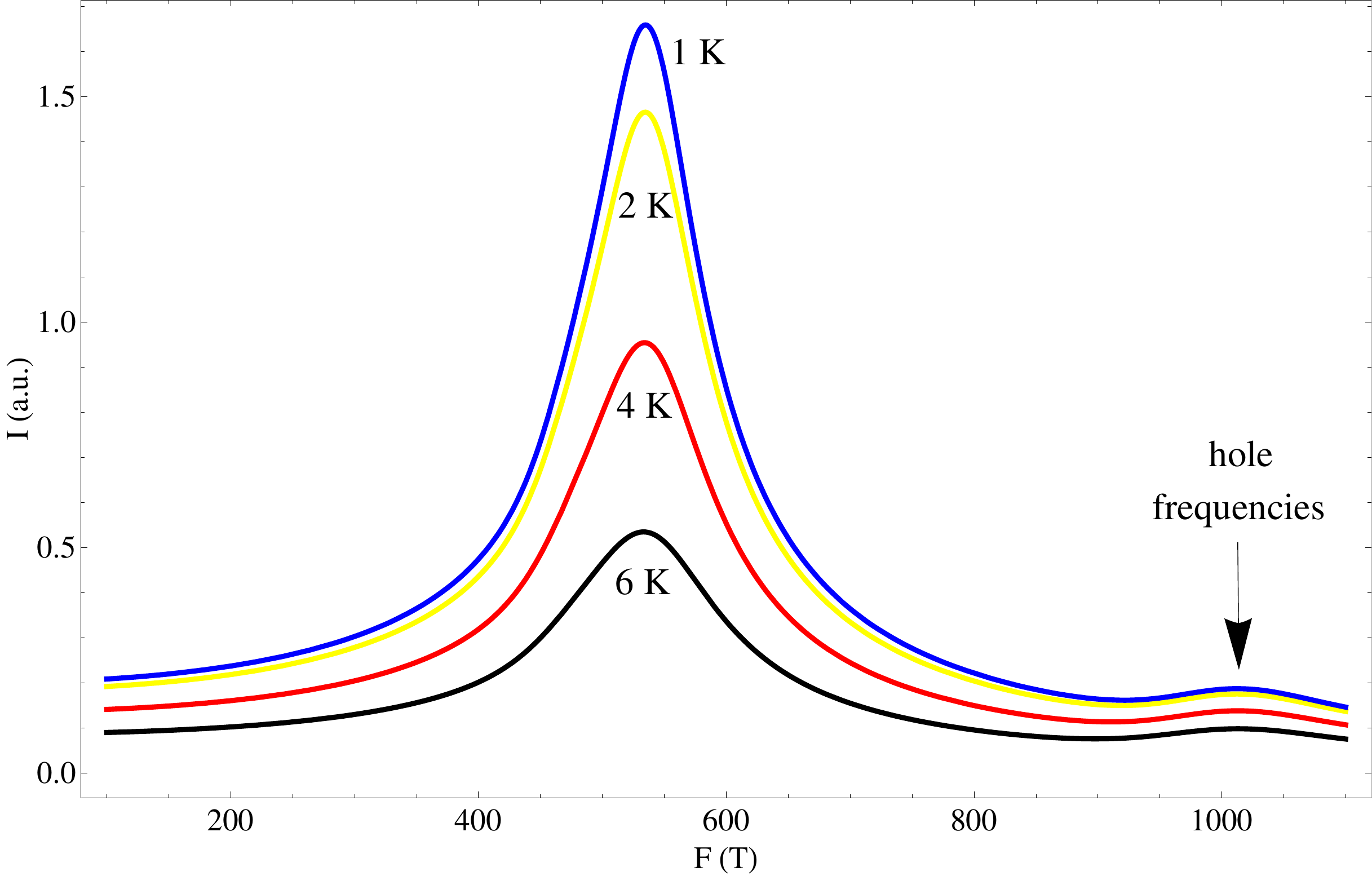}
\end{center}
\caption{Absolute value of the Fourier transform, $I$, in arbitrary units of the  oscillations of the magnetization as a function of the frequency $F$ in units of Tesla with no windowing  and  the field range of  $1/60 \le 1/H \le 1/20$.}
\label{fig:M1}
\end{figure}

Similarly, we also plot the oscillations of the magnetization as a function of $1/H$ in Fig.~\ref{fig:M2}, 
but it is difficult to detect multiple frequencies with naked eyes. Even in the Fourier transform over a range $1/60 \le 1/H \le 1/20$, shown in Fig~\ref{fig:M1}, the multiple electron pocket frequencies known to be present in the formula are not resolved.
The Fourier transform is now monotonic as a function of temperature unlike the results for the specific heat. The arrow indicates weak, unresolved  hole pocket  frequencies around $1000 \;\textrm{T}$; see Fig.~\ref{fig:Freqs2}. The two electron pocket frequencies at $535\;\textrm{T}$ and $538\;\textrm{T}$ strongly overlap and lead to a large amplitude. The spin interference factor $\cos \left(\pi m^{*}/m\right)$ play an important role in the respective weights of the various frequencies.

\section{Conclusion}
We have argued that bilayer splitting and warping of the electronic dispersion in $k_{z}$ are  necessary consequences of a reconstructed  Fermi liquid normal state, and measurements in a tilted magnetic field can be useful in probing the frequency spectra of quantum oscillations. The small value of the warping is intimately connected to the large ratio  of the in-plane to $c$-axis  resistivities. However, the magnitude of bilayer splitting necessary to produce overall consistency with experiments for the in-phase DDW order is strongly renormalized ($\sim 10$ meV) from the band structure value ($\sim 300$ meV). Note that the distance between the layers is only 3.25\AA, similar  to the in-plane lattice constant. In contrast, with the out-of-phase DDW order a larger value of bilayer splitting  ($\sim 37$ meV) can be tolerated. This is an important consequence of the out-of-phase DDW order.  Although strong electronic correlations in the underdoped regime must be responsible for such renomalized parameters, a convincing explanation is missing despite many speculations, especially because the effective mass is only about twice the free electron mass. It would be interesting to carry out these QO measurements for larger hole-doping for which we generally expect the splitting to increase, unless some other effects involving the decrease of the magnitude of the order parameter intervenes. It is worth emphasizing once again that even in heavily overdoped Bi2212, the renormalization of the observed bilayer splitting, 88 meV, in ARPES, as compared to the band structure value of 300 meV,  is still not understood.

The calculations presented here can be easily extended within a mean field theory  to SDW and incommensurate order along the lines discussed elsewhere~\cite{Dimov:2008}.  A more illuminating exercise is to compare and contrast quantum oscillations in hole and electron doped cuprates~\cite{Eun:2009}. The likely differences in the upper critical fields lead to important physical differences. Further  work along this direction is in progress. The triplet DDW~\cite{Nayak:2000} at the simplest mean field level  produces results similar to SDW, which is also a triplet order parameter, but with orbital angular momentum zero. Such triplet order parameters are necessary to explain the experiments~\cite{Sebastian:2009} involving the non-existence of spin zeros in QO. However, more recent experiments~\cite{Ramshaw:2010} have revealed spin zeros and have concluded that quasiparticles behave like charge-$e$, spin-$1/2$ fermions with a $g$-factor consistent with  2.2. This is strongly indicative of a singlet order parameter, but not a  triplet order in the particle-hole channel, such as SDW or triplet DDW~\cite{Garcia:2010}. 

Although we have obtained  consistency with experiments using Fermi liquid theory, it is not certain that non-Fermi liquid aspects should be ignored, at least insofar as underdoped YBCO is concerned. Convincing explanation of the lack of  the hole pocket frequencies required by the Luttinger sum rule~\cite{Luttinger:1960,Chubukov:1997,Dzyaloshinskii:2003} and the inconsistency with Fermi arcs observed in ARPES, albeit in zero magnetic field, are intriguing. We know of one  example, the $\nu=1/2$ quantum Hall effect, which despite being a non-Fermi liquid has a phenomenology similar to a Fermi liquid in many respects~\cite{Halperin:1993}. The situation in NCCO is clearer~\cite{Eun:2009}, however. We hope that our work will shed light on these exciting set of experimental developments.

\ack
We thank Cyril Proust for many interesting correspondence and S. Sebastian for drawing our attention to the measurements in a tilted magnetic field. We are grateful to D. Bonn and G. Boebinger for sharing with us their unpublished data. We also  thank E. Abrahams for a critical reading of the manuscript. This work is supported by NSF under the Grant DMR-0705092. DGA acknowledges financial support to Fundacion Cajamadrid through its
grant program and from the Spanish MCInn through the research project ref.
FIS2007-65702-C02-02.

\end{document}